\documentclass[aps, prl, showpacs, amsmath, twocolumn, groupedaddress]{revtex4}
\usepackage{graphics}
\usepackage{graphicx}
\usepackage{subfigure}
\usepackage{float}
\usepackage{longtable}
\usepackage{amssymb}
\usepackage{bm}

\begin{document}
\title{Symmetry of carrier-envelope phase difference effects in strong-field,
few-cycle ionization of atoms and molecules}
\author{Christian Per Juul Martiny}
\author{Lars Bojer Madsen}
\affiliation{Department of Physics and Astronomy,
  University of  Aarhus, 8000 {\AA}rhus C, Denmark}

\date{\today}

\begin{abstract}
In few-cycle pulses, the exact value of the carrier-envelope phase
difference (CEPD) has a pronounced influence on the ionization
dynamics of atoms and molecules. We show that for atoms in
circularly polarized light, a change in the CEPD is mapped uniquely
to an overall rotation of the system, and results for arbitrary CEPD
are obtained by rotation of the results from a single calculation
with fixed CEPD. For molecules this is true only for linear
molecules aligned parallel with the propagation direction of the
field. The effects of CEPD are classified as geometric or
non-geometric. The observations are exemplified by strong-field
calculations on hydrogen.
\end{abstract}

%32.80.Rm Multiphoton ionization and excitation to highly excited states (e.g., Rydberg states)

\maketitle

\begin{figure}
\begin{center}
    \includegraphics[width=0.45\columnwidth]{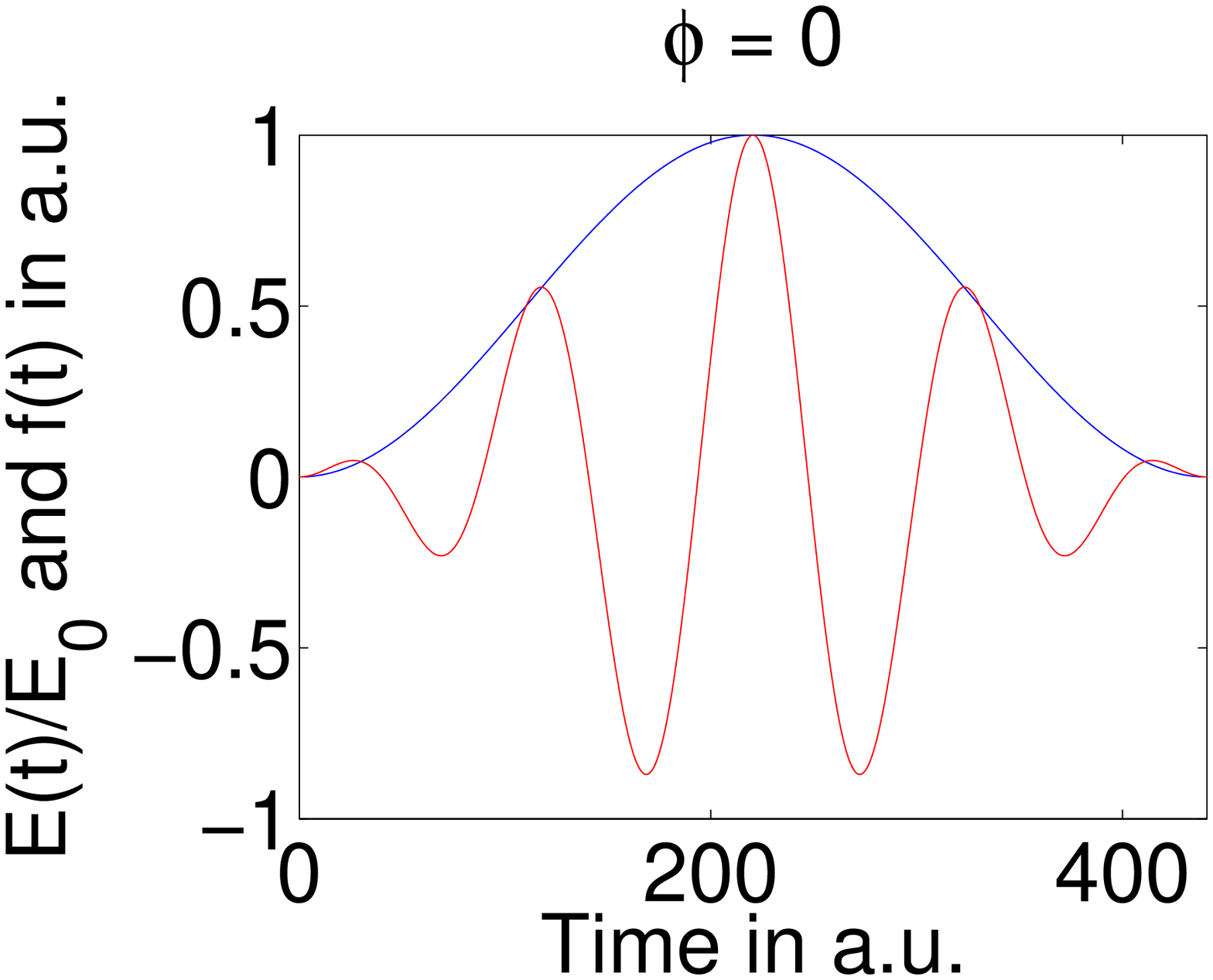}
    \includegraphics[width=0.45\columnwidth]{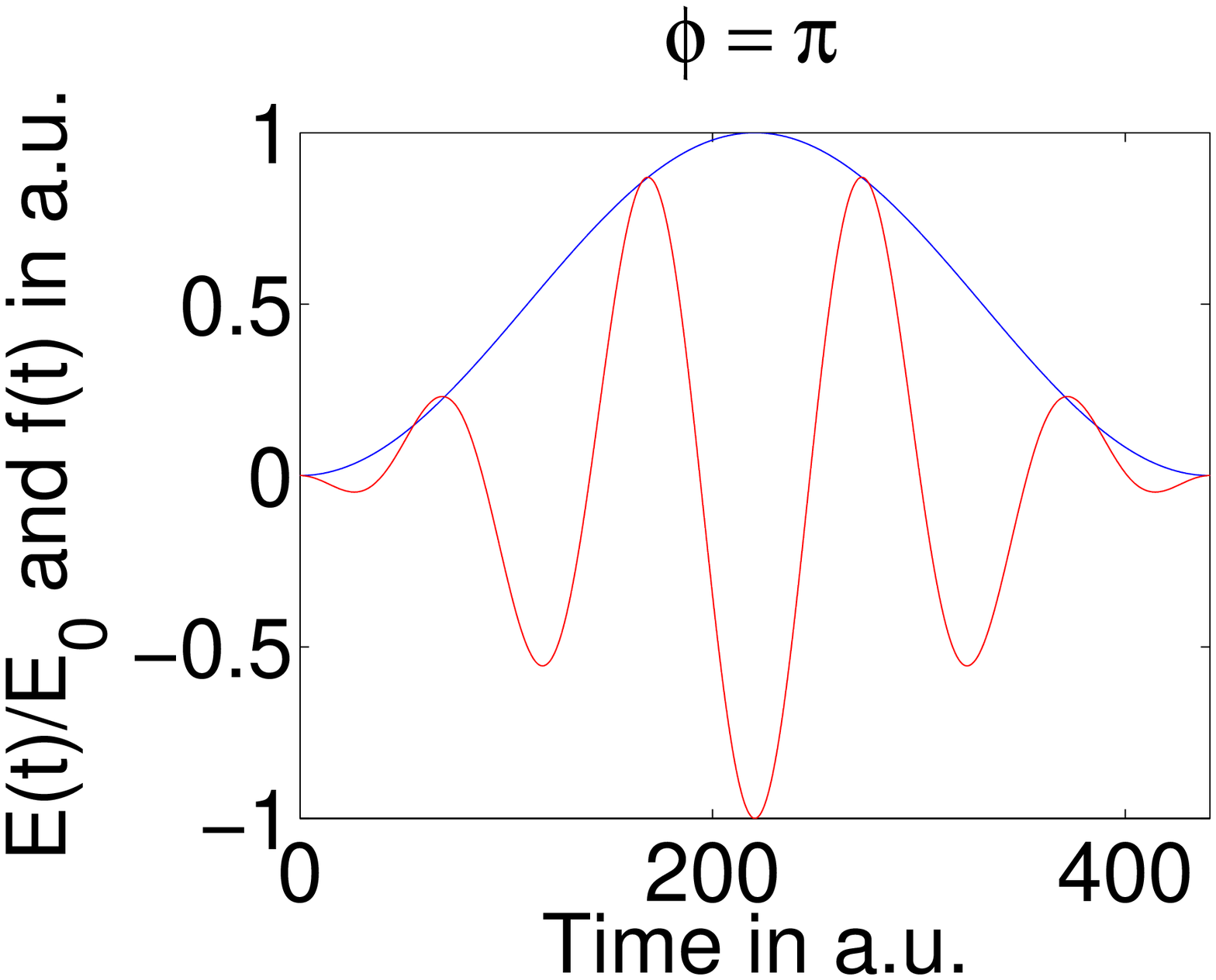}
    \caption{(Color online). The electric field $E(t)/E_{0}$,  normalized to the peak field
    strength $E_0$ and the pulse envelope $f(t)$ as a function
     of time for two values of the CEPD, $\phi$.  The electric field points in opposite directions for
$\phi=0$ and $\phi=\pi$.
In ionization, the electric field will shake the electron until it
gains enough energy to escape the Coulomb potential, and the angular
distribution will depend on CEPD because the electric field (and the
force $\vec{F}=-\vec{E}$) points in opposite directions for $\phi=0$
and $\phi=\pi$.
     The field parameters are $\tau=441 \text{ a.u.}$ and $\omega=0.057 \text{ a.u.}$ (800 nm).}
     \label{fig:fig1}
\end{center}
\end{figure}

Nowadays,  it is possible to construct and control intense laser
pulses with only a few optical cycles~\cite{Brabec00}, i.e., pulses
described by a vector potential of the form $\vec{A}(t)=A_{0}
f(t)\sin(\omega(t-\frac{\tau}{2})+\phi)\hat{e}$, where $A_{0}$ is
the amplitude, $f(t)$ is the envelope, $\tau$ is the pulse length,
$\omega$ is the frequency, $\phi$ is the carrier-envelope phase
difference (CEPD), and $\hat{e}$ is the polarization vector. The
corresponding electric field is obtained from the vector potential
by $\vec{E}(t) = - \partial_t {\vec A} (t)$, and is shown in
Fig.~\ref{fig:fig1} for a $\sin^2$ envelope. Such pulses can be used
to probe molecular and atomic dynamics on a very short time scale
\cite{Zewail, Attophysics}. The associated ionization dynamics
becomes sensitive to the exact shape of the pulse and the
carrier-envelope phase difference (CEPD) (see Fig.~\ref{fig:fig1}).
This dependence may be understood by the exponential dependence of
the ionization rate on the instantaneous field strength and the
corresponding emergence of the electron into the field-dressed
continuum at specific instants of time during the
pulse~\cite{Scrinzi06}. Asymmetries in the photoelectron spectrum
may give information about the CEPD, and hence help in the
characterization of the field. Electrons released at different times
are accelerated to different final momenta and this fact is
exploited in attosecond streaking~\cite{Itatani02} to map the time
distribution of the pulse into a momentum distribution of the
photoelectron, and to characterize the ultra-fast
pulses~\cite{Bandrauk02}. Hence, just as few-cycle pulses are
diagnostic tools for  atoms and small molecules, the very same
systems serve as diagnostic tools for the pulses
themselves~\cite{Goulielmakis04}. The latter statement, of course,
assumes that an accurate theoretical description is at hand for the
pulsed laser--matter interaction. It is the purpose of this work to
add further to this understanding. In particular we are concerned
with the CEPD effects, and a geometric interpretation of these.

Carrier-envelope phase difference effects were studied theoretically
with emphasis on CEPD-induced spatial asymmetries in the
ATI-spectrum/angular distribution in a number of papers on
strong-field ionization of
atoms~\cite{Cormier98,Dietrich00,Hansen01,Bandrauk,Becker} and
molecules~\cite{Collins}, strong-field dissociation~\cite{Roudnev},
and high-harmonic generation~\cite{Highharmonic}. The asymmetries
can be used to extract information about and ultimately to measure
the CEPD. In Ref.~\cite{Silvestri} a spatial asymmetry in ionization
with few-cycle circular polarized laser pulses was observed for the
first time. In Ref.~\cite{Krausz} the generation of intense
few-cycle laser pulses with stable CEPD was demonstrated, and a way
of measuring CEPD, based on soft-X-ray radiation was presented. In
Ref.~\cite{Paulus} a spatial asymmetry in the ionization with
few-cycle linear polarized laser pulses was measured, and the CEPD
was determined with an estimated error of $\pi/10$. Asymmetries in
the ionization signal combined with an attosecond pump
pulse~\cite{Bandrauk02} were used to measure directly the field of a
linearly polarized few-cycle pulse~\cite{Goulielmakis04}. Very
recently, the detailed control over the CEPD played a crucial role
in attosecond electron dynamics~\cite{Paulus,Sola,Remetter}, and
experiments supporting the present findings  have been reported at
conferences~\cite{Schlup}.

In this work, we present a systematic theoretical study of the
symmetry of the response of atomic and molecular systems under a
change of the CEPD in few-cycle pulses, and we exemplify the
discussion with a study of the differential electron momentum
distribution for hydrogen under such pulses.

We start out by considering an $n$-electron atom interacting with a
few-cycle circularly polarized laser pulse described by the vector
potential
\begin{align}
\label{A-felt}
\vec{A}(\phi,t,\vec{r})=&\frac{A_{0}}{\sqrt{2}}f(\eta) \\
&\times\left(\cos(\eta+\phi+\frac{\pi}{2})\vec{e}_{x}+\sin(\eta+\phi+\frac{\pi}{2})\vec{e}_{y}
\right),\nonumber
\end{align}
with $f(\eta)=\sin^{2}(\frac{\eta}{2N})$ the envelope, $N$  the
number of optical cycles, $\eta=\omega t-kz$, $\omega$ the
frequency, and ${\vec k } = k {\vec e}_z$ the wave vector. In the
present case, with full inclusion of the spatial dependence of the
field, the interaction of the atom with the field is obtained by the
minimal coupling ${\hat{\vec p}} \rightarrow {\hat{\vec p}} +
\vec{A}$ and the time-dependent Schr\"{o}dinger equation reads
[atomic units (a.u.) with $m_e=e=a_0=\hbar=1$ are used throughout],
\begin{equation}
\label{eq:schr} i\frac{\partial}{\partial t}\Psi=\left(
H_{0}+\sum^{n}_{j=1}\vec{A}(\phi,t,{\vec r}_{j})\cdot\hat{{\vec
p}}_{j}+\frac{nA^{2}_{0}}{4}f^{2}(\eta)\right)\Psi,
\end{equation}
where $H_{0}$ is the field-free Hamiltonian. We are interested in
relating this equation to an equation for the $\phi=0$ case. To this
end, we note that the unitary operator
$D(\phi)=\exp\left(iJ_{z}\phi\right)$, where $J_{z}$ is the total
angular momentum, corresponds to a rotation of our system by an
angle $-\phi$ around the $z$-axis. Since $H_{0}$ is invariant under
rotations around the $z$-axis, the transformed wave function
$\Psi'=D(\phi)\Psi$ satisfies the Shr\"{o}dinger equation,
\begin{align}
i\frac{\partial}{\partial
t}\Psi'=&(H_{0}+\sum^{n}_{j=1}D(\phi)\vec{A}(\phi,t,{\vec r}_{j})\cdot\hat{\vec{p}}_{j}D^{\dagger}(\phi)\\
&+\frac{nA^{2}_{0}}{4}D(\phi)f^{2}(\eta)D^{\dagger}(\phi))\Psi'\nonumber,
\end{align}
and using the Baker-Hausdorff lemma~\cite{Sakurai} we obtain,
\begin{equation}
\label{eq:trans_schr}
 i\frac{\partial}{\partial
t}\Psi'=\left(H_{0}+\sum^{n}_{j=1}\vec{A}(0,t,{\vec r}_{j})\cdot
\hat{{\vec{p}}}_{j}+ \frac{nA^{2}_{0}}{4}f^{2}(\eta)\right)\Psi'.
\end{equation}
When we compare \eqref{eq:schr} and \eqref{eq:trans_schr}, we see
that a change in the CEPD from $\phi=0$ to $\phi=\phi\hspace{1mm}'$
corresponds to a rotation of our system around the $z$-axis by the
angle $\phi\hspace{1mm}'$. This fact is not only theoretically
interesting, but also helpful in practical calculations since this
symmetry property reduces the number of computations one has to
perform to a single one -- all other results are obtained by
suitable rotations. For instance, imagine we are interested in the
differential ionization probability
$\frac{dP_{fi}}{dq_{x}dq_{y}}(\phi)$ for the momenta $q_x, q_y$ in
the polarization plane and for a general $\phi>0$. Then we calculate
$\frac{dP_{fi}}{dq_{x}dq_{y}}(\phi=0)$, and rotate the result
counterclockwise by an angle $\phi$ to obtain
$\frac{dP_{fi}}{dq_{x}dq_{y}}(\phi)$.

In the above derivation it is essential that the field-free
Hamiltonian $H_{0}$ is invariant to rotations around the $z$-axis.
If this is not the case, the proof breaks down. As an example we
look at ionization of diatomic molecules, or more generally linear
molecules, in circularly polarized few-cycle laser pulses. If the
molecule is aligned along the laser propagation direction, then the
field-free Hamiltonian still has the required symmetry and the
theorem holds. If the molecule is not aligned along this axis, the
system does not have the required symmetry and CEPD effects can not
be reduced to a geometrical rotation. Accordingly, we may make a
distinction between rotational invariant atomic and molecular
systems where the CEPD effects are purely geometric rotations, and
systems which are not rotational invariant in which case true {\it
non-geometrical} CEPD effects occur. For example, the results on
strong-field ionization of K$_2^+$ with the molecule in the
polarization plane~\cite{Collins} belong to the latter category.

As an illustration of the present findings, we consider ionization
of atomic hydrogen, H(1s), in the strong-field approximation
(SFA)~\cite{Keldysh}. We assume the dipole approximation which means
that $\eta = \omega t$ in \eqref{A-felt} and use that $\vec{E} = -
\partial_t \vec{A}$. The probability amplitude for direct
ionization reads
\begin{equation}
\label{eq:T}
T_{fi}=-i\int^{\tau}_{0}\langle\Psi_{f}(\vec{r},t)|\vec{E}
\cdot {\vec r} |\Psi_{i}(\vec{r},t)\rangle dt,
\end{equation}
where $\Psi_{i}(\vec{r},t)$ is the 1s ground state wave function.
The final state $\Psi_{f}(\vec{r},t)$ is represented by a
Coulomb-Volkov wave function~\cite{CoulombVolkov} with asymptotic
momentum $\vec{q}$. The integral in (\ref{eq:T}) is analyzed as
in~\cite{CoulombVolkov} and the numerical evaluation is performed
with Gauss-Legendre quadrature.
%$\Psi_{f}(\vec{r},t)=\psi_{f}(\vec{r},t)L(\vec{r},t)$ where
%$\psi_{f}(\vec{r},t)=\varphi(\vec{r})\exp(-i\epsilon_{f}t)$ is the
%incoming regular Coulomb wave function, $L$ is the field dependent
%part of a Volkov wave function,
%$L(\vec{r},t)=\exp\left(i\vec{A}\cdot\vec{r}-i\vec{q}\cdot\int^{t}_{0}\vec{A}(t')dt'-\frac{i}{2}\int^{t}_{0}A^{2}(t')dt'\right)
%$, $\vec{q}$ is the momentum and $\epsilon_{f}$ is the energy of the
%ejected electron. For vanishing field, the Coulomb-Volkov wave
%function is reduced to an incoming Coulomb wave. We
%follow~\cite{CoulombVolkov} and define the functions,
%$h(t)=i\left(\frac{q^{2}}{2}-\epsilon_{i}\right)+i\vec{q}\cdot\vec{A}(t)+\frac{i}{2}A^{2}(t)$,
%$f(t)=\exp\left(\int^{t}_{0}h(t')dt'\right)$, and
%$g(t)=\int\psi^{\ast}_{f}(\vec{r})\exp\left(-i\vec{A}(t)\cdot\vec{r}
%\right)\psi_{i}(\vec{r})d\vec{r}.$ The function $h$ describes the
%energy of the ejected electron, while $f$ comes from the time
%dependent part of the Coulomb-Volkov wave. The functions $h$, $f$,
%and $g$ may be evaluated analytically~\cite{Nordsieck}. Using that
%$\vec{A}(\tau)=\vec{A}(0)=\vec{0}$, we obtain the probability
%amplitude $T_{fi}=\int^{\tau}_{0}h(t)g(t)f(t)dt$ which we evaluate
%numerically using Gauss-Legendre quadrature.
By exploiting the
invariance of the dot-product under rotations ($\vec{E}(\phi) \cdot
\vec{r} = [R_z(\phi) \vec{E}(\phi=0)]\cdot [R_z(\phi) R_z(-\phi)
\vec{r}] = \vec{E} (\phi=0) \cdot \vec{r}\,'$, where $R_z(\phi)$ is
the $3\times 3$ matrix describing rotations around the $z$ axis),
the SFA is readily shown to respond to CEPD changes like the exact
theory discussed above.

\begin{figure}
\begin{center}
    \includegraphics[width=0.45\columnwidth]{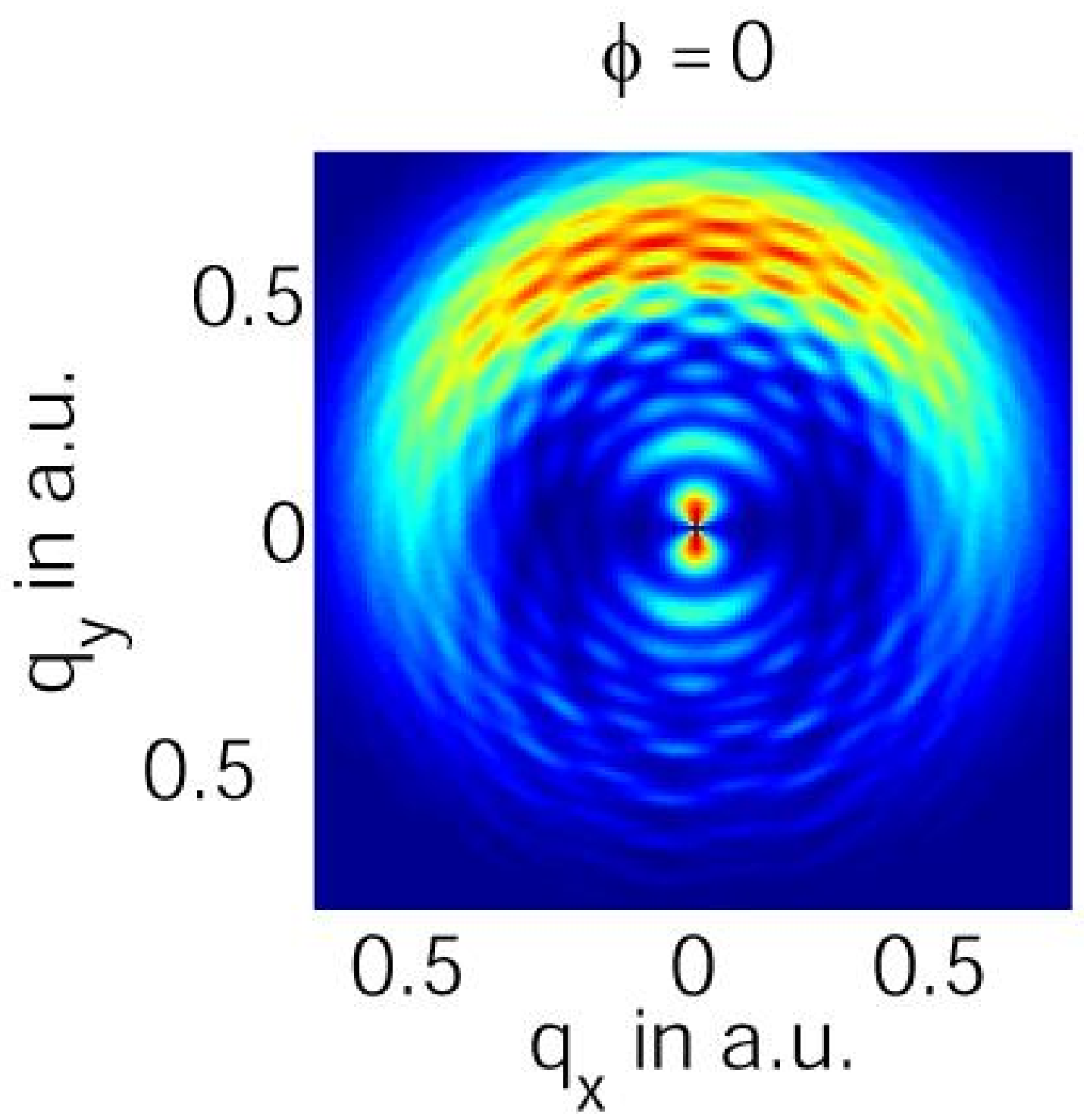}
    \includegraphics[width=0.45\columnwidth]{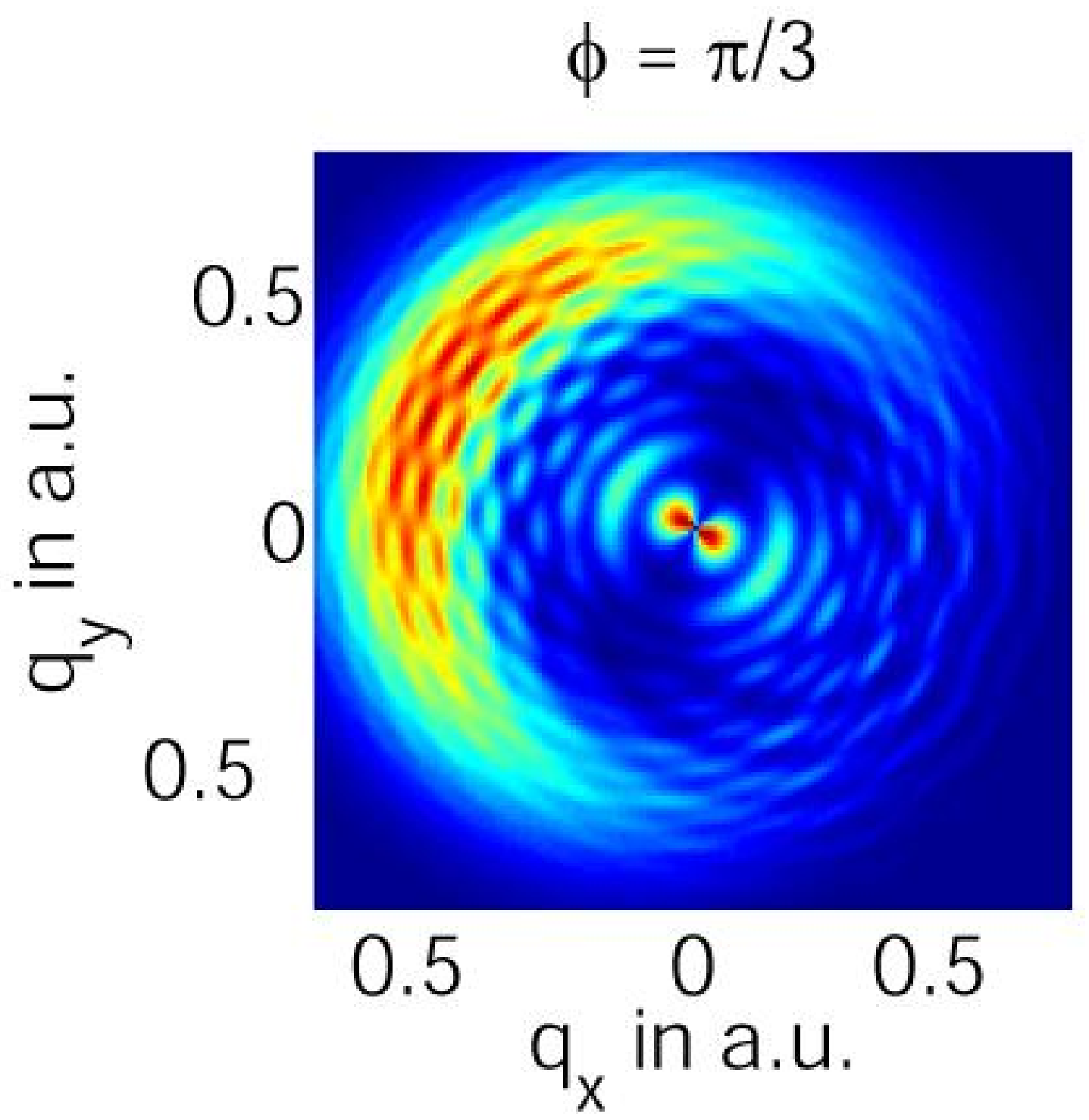}
    \includegraphics[width=0.45\columnwidth]{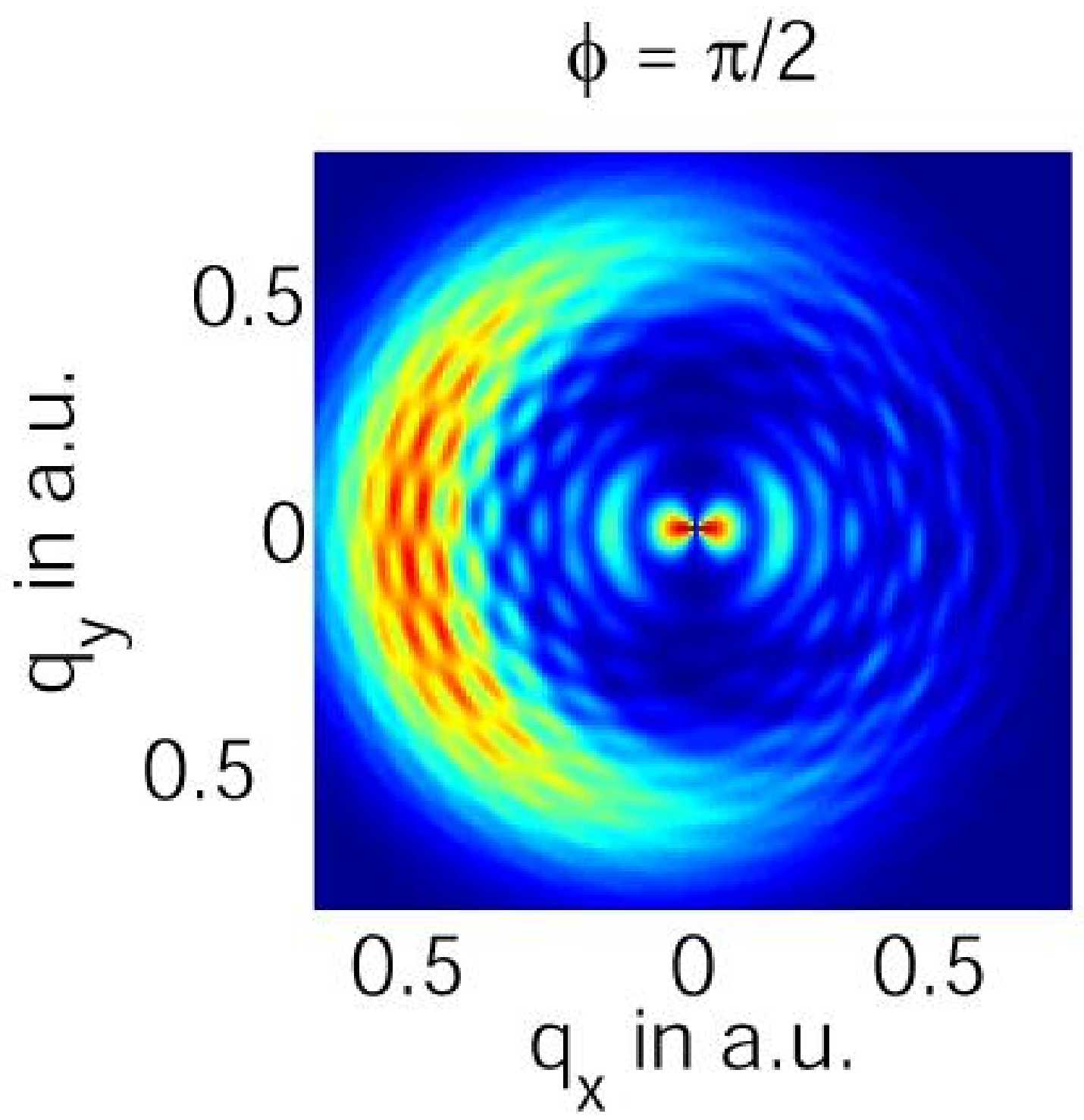}
    \includegraphics[width=0.45\columnwidth]{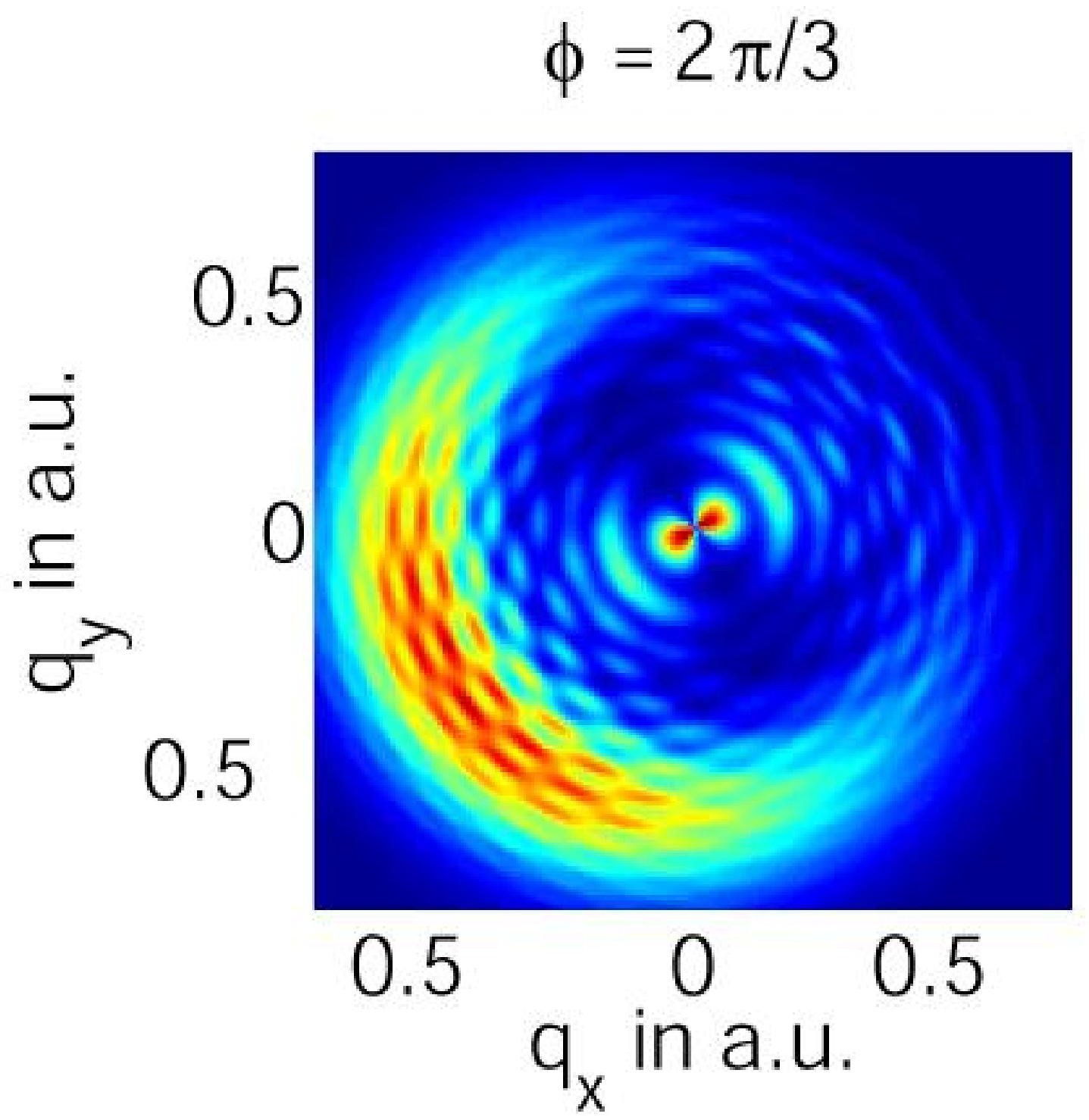}
    \caption{(Color online) The $(q_{x},q_{y})$ distribution for strong-field ionization
     of H(1s) for various values of $\phi$, with
     $I=5.0 \times 10^{13}$\,W/cm$^2$, $\omega = 0.057$
     corresponding to a central wavelength of
     $800$\text{nm} and three cycles, $N=3$.
     The grid size is $dq_{x}=dq_{y}=0.01$.}\label{fig:fig2}
\end{center}
\end{figure}

Once $T_{fi}$ is known, a simple numerical integration over $q_{z}$
gives us the $(q_{x},q_{y})$ distribution$
\frac{dP_{fi}}{dq_{x}dq_{y}}=\int |T_{fi}|^{2} \, dq_z.$ Figure
\ref{fig:fig2} presents the calculated distribution for various
values of $\phi$.
For varying $\phi$, the distribution rotates in accordance with the
general theory. For $\phi=0$, $\frac{dP_{fi}}{dq_{x}dq_{y}}$ has a
peak around $(q_{x}=0,q_{y}\sim0.6)$ (this peak corresponds to about
14 photons above threshold), and is almost symmetric around the line
$q_{x}=0$. As $\phi$ increases, this line is turned
counterclockwise, so that $\frac{dP_{fi}}{dq_{x}dq_{y}}$ has a peak
around $(q_{y}=0,q_{x}\sim-0.6)$ for $\phi=\frac{\pi}{2}$.

\begin{figure}
\begin{center}
    \includegraphics[width=0.45\columnwidth]{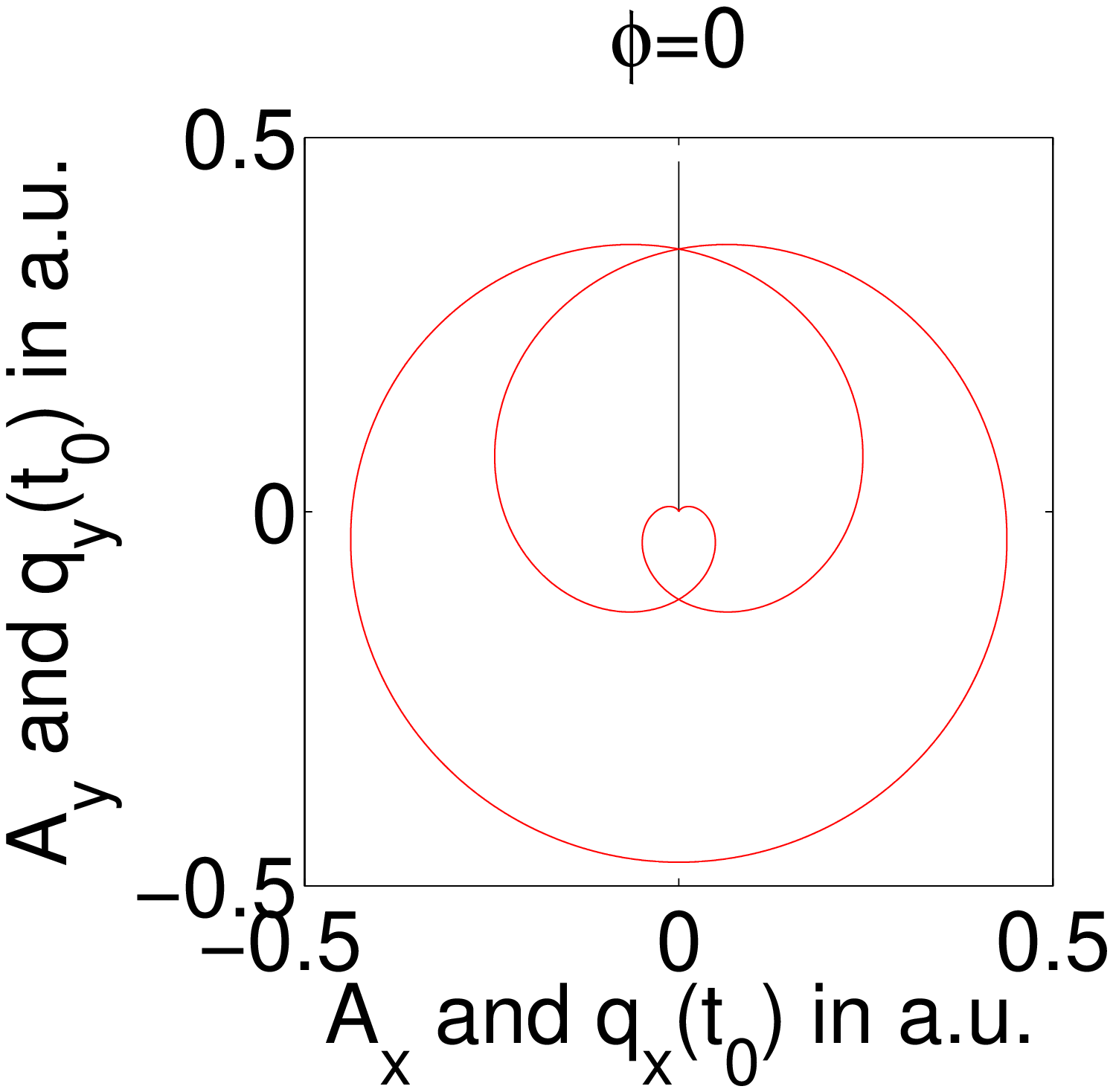}
    \includegraphics[width=0.45\columnwidth]{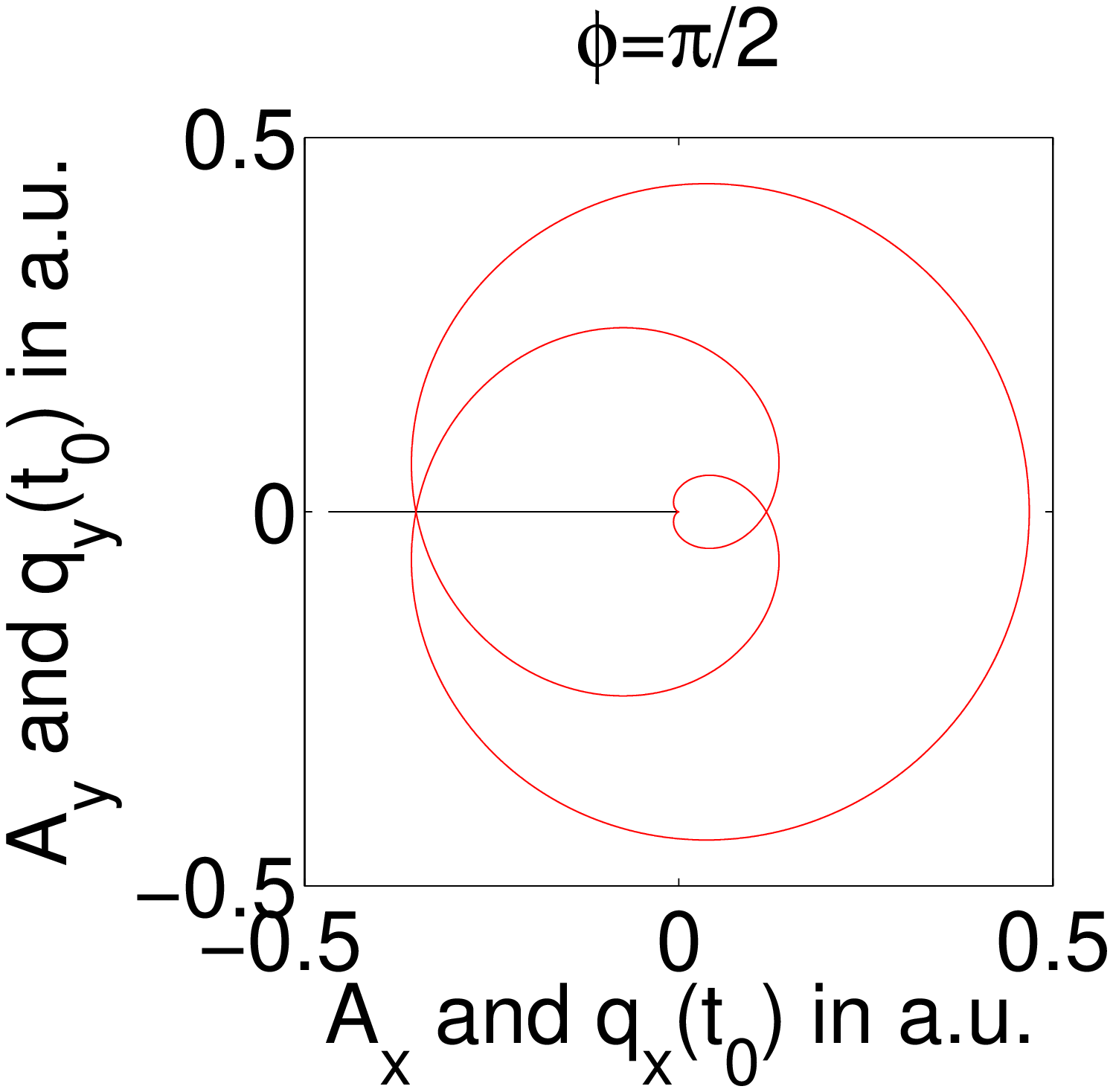}
    \caption{The curves show the time-dependent vector potential and the lines show $\vec{q}^{d}$
    for $\phi=0,\pi/2$.
    Laser parameters are as in
    Fig. \ref{fig:fig2}.}\label{fig:classical}
\end{center}
\end{figure}

Some of the features can be explained by a semiclassical two-step
model~\cite{Bandrauk}: first the electron escapes to the continuum
at $t=t_{0}$  with velocity $\vec{v}(t_{0})=\vec{0}$. Second, it
moves like a classical particle under the influence of the external
Coulomb and laser fields. If, we neglect the Coulomb potential,
which is justified if the field is very strong, Newton's second law
tells us that $q_{x}(t)=A_{x}(t)+q^{d}_{x}$,
$q_{y}(t)=A_{y}(t)+q^{d}_{y}$ where $q^{d}_{x}=-A_{x}(t_{0})$ and
$q^{d}_{y}=-A_{y}(t_{0})$. This means that the photoelectron
direction, after the pulse where $\vec{A}$ vanishes, is determined
by the vector $\vec{q}^{\hspace{1mm}d}=(q^{d}_{x},q^{d}_{y})$. For
simplicity $t_{0}$ is chosen to be $\tau/2$, which is the time when
$|\vec{E}(t)|$ reaches its maximum. Figure \ref{fig:classical} shows
plots of the time-dependent vector potential and
$\vec{q}^{\hspace{1mm}d}= -\vec{A}(t_0)$ for various values of
$\phi$.
The model explains how the preferred direction(the line of symmetry)
of the photoelectron depends on CEPD.

Finally we consider the case of a linearly polarized laser field
and return to the question of symmetry. The vector potential in
the linear case, $\vec{A}(\phi,t,\vec{r})=A_0
f(\eta)\sin(\eta+\phi+\pi/2)\hat{e}_{y}$, can be written as
$\vec{A}(\phi,t,\vec{r})=\frac{1}{\sqrt{2}}(\vec{A}_{L}(\phi,t)-\vec{A}_{R}(\phi,t))$,
where
$\vec{A}_{L/R}(\phi,t,\vec{r})=\frac{A_0}{\sqrt{2}}f(\eta)(\cos(\eta+\phi+\pi/2)\hat{e}_{x}
\pm \sin(\eta+\phi+\pi/2)\hat{e}_{y})$ describe left$(+)$ and
right$(-)$ hand circularly polarized fields. This relation
suggests that we can characterize CEPD effects for a linearly
polarized pulse similarly to circularly polarized pulses.
Unfortunately, in exact theory, it is not possible to separate the
effects of the circularly polarized fields, since they do not
individually commute with the field-free Hamiltonian. \\Figure 4
shows two examples for a linear field corresponding to the
$\phi=0$ and $\phi=\pi/2$ cases. These results may be interpreted
in terms of the classical model discussed above ($q_y=-A_y(t_0)$).
For $\phi=0$, the vector potential peaks in the negative $y$
direction at the peak of the envelope and produces electrons with
positive momentum, $q_{y}=-A_{y}(\frac{\tau}{2})$. For $\phi =
\pi/2$, the vector potential is antisymmetric with respect to the
peak of the envelope, and the electron spectrum reflects this in a
forward-backward symmetry. For $\phi=\pi/2$, the spectrum has
moved to smaller momenta since the two maxima of the vector
potential are lower in this case. Notice, however, that the model
can not explain the size of the shift. There are two reasons for
this discrepancy. First, the model neglects the Coulomb potential.
Second we assume that $v_{0}=0$, which is not necessarily correct.
\begin{figure}
\begin{center}
    \includegraphics[width=0.45\columnwidth]{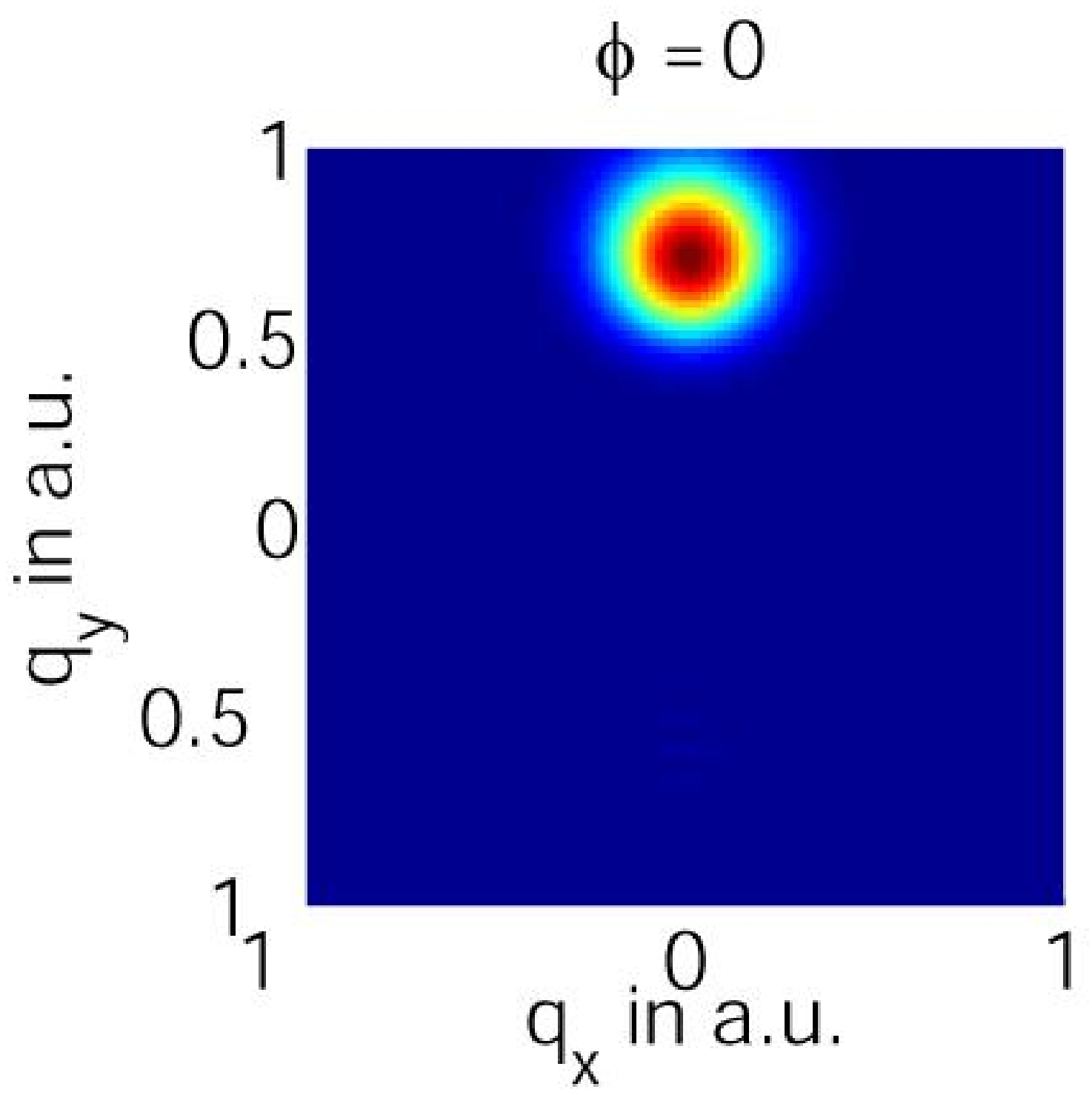}
    \includegraphics[width=0.45\columnwidth]{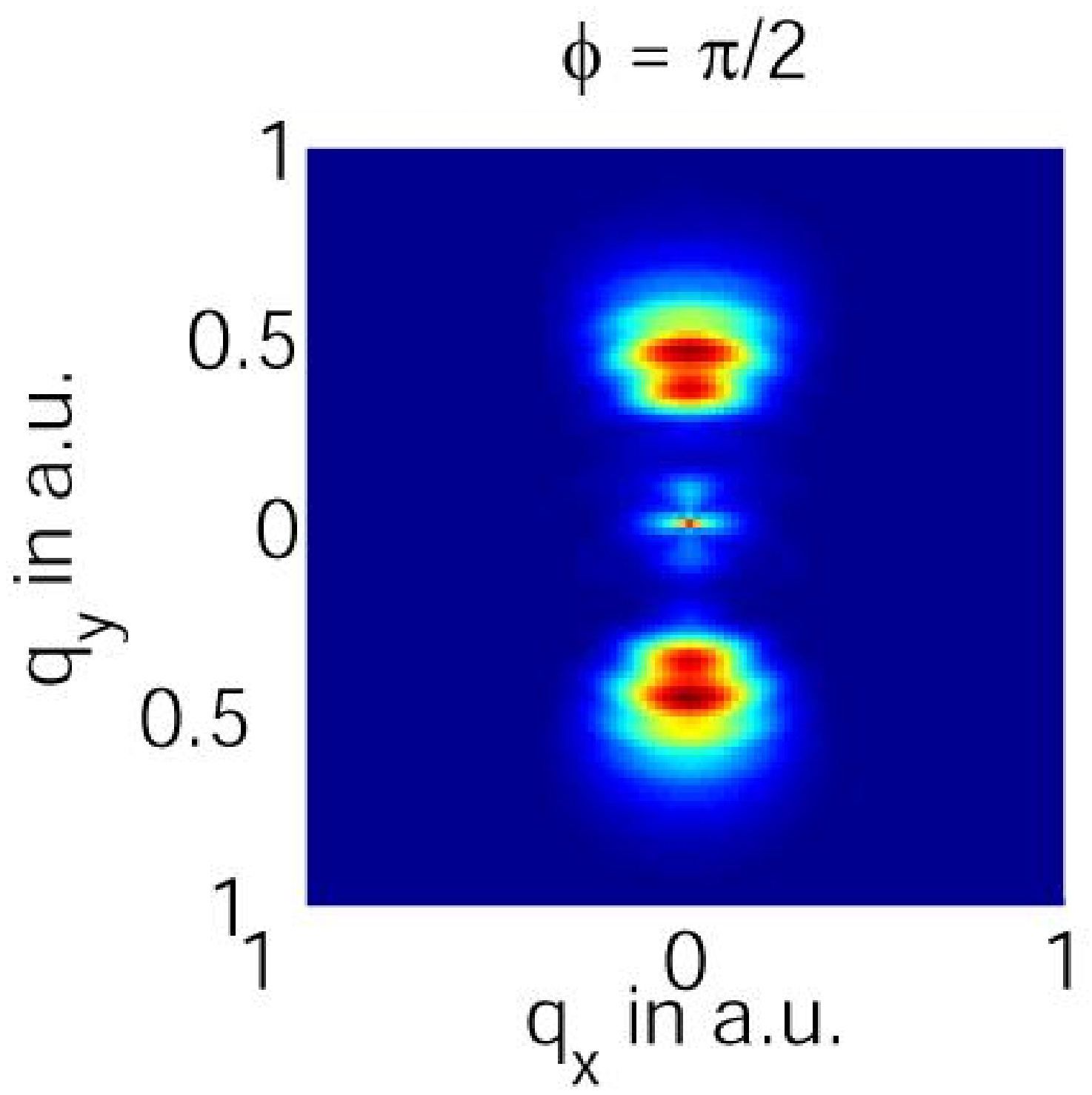}
    \caption{(Color online) The $(q_{x},q_{y})$ distribution for strong-field ionization
     of H(1s) by a linearly polarized field for $\phi=0,\pi/2$, with
     $I=5.0 \times 10^{13}$\,W/cm$^2$, $\omega = 0.057$
     corresponding to a central wavelength of
     $800$\text{nm} and three cycles, $N=3$.
     The grid size is $dq_{x}=dq_{y}=0.02$.}\label{fig:fig4}
\end{center}
\end{figure}

In this work, we provided an exact characterization of CEPD
effects in the interaction between circularly polarized few-cycle
pulses and multielectron atoms and molecules including full
account of the spatial dependence of the field (non-dipole
effects). For systems which are invariant to rotations around the
propagation direction of the field, a change of the CEPD
corresponds to a rotation of the systems. For rotationally
non-invariant systems, true non-geometrical CEPD effects may be
observed. Calculations on H(1s) quantified the effects, and showed
that observation of the momentum distribution can be used to
measure CEPD for a circularly polarized laser pulse. For linearly
polarized fields, the electrons spectrum was interpreted in terms
of a classical model.

There is currently considerable interest in control of the response
of atoms and small molecules to a strong few-cycle
pulse~\cite{Scrinzi06}, e.g., in connection with alignment dependent
high-harmonic generation from small
molecules~\cite{Kanai05,Itatani05}. Here the harmonics provide a
delicate probe of the molecular wave packet dynamics and, in turn,
steering the electronic wave packets controls the harmonics. The
carrier-envelope-phase difference is one of the essential control
parameters in few-cycle pulse and attosecond science.

\begin{acknowledgments}
We thank T.K. Kjeldsen and U.V. Poulsen for useful discussion.
L.B.M. was supported by the Danish Natural Science Research Council
(Grant No. 21-03-0163) and the Danish Research Agency (Grant. No.
2117-05-0081).
\end{acknowledgments}

%\bibliography{bibfile}
\end{document}